\documentclass[pra,aps,twocolumn,floatfix,showpacs]{revtex4}
\usepackage{graphics}
\begin{document}
\title{Quantum phases of Fermi-Fermi mixtures in optical lattices}
\author{M. Iskin$^1$ and C. A. R. S{\'a} de Melo$^{1,2}$}
\affiliation{$^1$Joint Quantum Institute, National Institute of Standards and Technology 
and University of Maryland, Gaithersburg, Maryland 20899-8423, USA. \\
$^2$School of Physics, Georgia Institute of Technology, Atlanta, Georgia 30332-0432, USA.}
\date{\today}

\begin{abstract}
The ground state phase diagram of Fermi-Fermi mixtures in optical lattices 
is analyzed as a function of interaction strength, population imbalance, filling fraction and 
tunneling parameters. It is shown that population imbalanced Fermi-Fermi mixtures reduce to strongly
interacting Bose-Fermi mixtures in the molecular limit, in sharp contrast to homogeneous or 
harmonically trapped systems where the resulting Bose-Fermi mixture is weakly interacting.
Furthermore, insulating phases are found in optical lattices of Fermi-Fermi mixtures 
in addition to the standard phase-separated or coexisting superfluid/excess fermion phases found 
in homogeneous systems. The insulating states can be a molecular Bose-Mott insulator (BMI), 
a Fermi-Pauli insulator (FPI), a phase-separated BMI/FPI mixture or a Bose-Fermi checkerboard (BFC). 

\end{abstract}
\maketitle

\section{Introduction}
\label{sec:introduction}

In recent experiments, superfluidity of ultracold $^6$Li Fermi atoms with population
imbalance was investigated in gaussian traps
from the Bardeen-Cooper-Schrieffer (BCS) to the Bose-Einstein condensation (BEC) limit~\cite{mit, rice, mit-2, rice-2}.
In contrast to the crossover physics found in population balanced systems~\cite{leggett, nsr, carlos, jan}, 
a phase transition from superfluid to normal phase as well as phase separation were observed~\cite{bedaque, pao, sheehy}.
More recently, experimental evidence for superfluid and insulating phases has also been reported
with ultracold $^6$Li atoms in optical lattices without population imbalance~\cite{mit-lattice}, 
after overcoming some earlier difficulties of producing Fermi 
superfluids from an atomic Fermi gas or from molecules of Fermi 
atoms in optical lattices~\cite{inguscio, modugno, kohl, stoferle, bongs}.

Before and after experiments, the possibility of interesting phases in population 
imbalanced ultracold fermions have attracted intense theoretical 
interest~\cite{bedaque, pao, sheehy, torma, pieri, yi, silva, haque, iskin-mixture, lobo, liu-mixture, mizushima, parish, iskin-mixture2}.
These works were focused on homogeneous or 
harmonically trapped Fermi superfluids, where the number of external control parameters is limited. 
For instance, in addition to population imbalance $P$ and scattering parameter $a_s$, several other 
experimental parameters can be controlled in optical lattices
such as the tunneling matrix element $t_\sigma$ between adjacent lattice sites, 
the on-site atom-atom interactions $g$, the filling fraction $F$, 
the lattice dimensionality $\cal{D}$ and the tunneling anisotropy $\eta = t_\downarrow/t_\uparrow$~\cite{bloch-review}. 
Therefore, optical lattices may permit a systematic investigation of phase diagrams 
and correlation effects in Fermi systems as a function of $g$, $P$, $F$, ${\cal D}$ 
and $\eta$, which has not been possible in any other atomic, nuclear and/or condensed matter systems. 

Arguably, one-species or two-species fermions loaded into optical lattices are
one of the next frontiers in ultracold atom research because of its greater tunability. 
The problem of one-species population balanced fermions in lattices has already been 
studied in the BCS and BEC limits~\cite{nsr, micnas, iskin-prb}, while recent works on population 
and/or mass imbalanced Fermi-Fermi mixtures in lattices were limited
only to the BCS regime~\cite{liu-lattice, torma-lattice, chen}.
In this manuscript, we extend our previous work describing balanced fermions and their s-wave and p-wave 
superfluid phases in optical lattices~\cite{iskin-prb}, and extend our recent Letter~\cite{iskin-lattice}
describing superfluid and insulating phases of Fermi-Fermi mixtures in optical lattices.
We discuss specifically the population balanced and imbalanced Fermi-Fermi mixtures of one-species 
(e.g. $^6$Li or $^{40}$K only) and of two-species (e.g. $^6$Li and $^{40}$K;
$^6$Li and $^{87}$Sr; or $^{40}$K and $^{87}$Sr),
as a function of $g$, $P$ and $F$ for fixed values of $\eta$ in optical lattices and throughout 
the BCS to BEC evolution. We also pay special attention to the emergence of insulating phases
in the strong attraction (molecular) limit. The existence of insulating phases in optical lattices
should be contrasted with its absence in homogeneous or 
harmonically trapped systems~\cite{pao, sheehy, iskin-mixture, iskin-mixture2}.
Furthermore, the present extention to two-species mixtures is timely due to very 
recent experimental work on $^6$Li and $^{40}$K mixtures~\cite{dieckmann-lik,grimm-lik} 
in harmonic traps, which have opened the possibility of trapping these mixtures 
in optical lattices as well.

Our main results are as follows.
Using an attractive Fermi-Hubbard Hamiltonian to describe one-species or 
two-species mixtures, we obtain the ground state phase 
diagram containing normal, phase-separated and coexisting superfluid/excess fermions, and 
insulating regions.
We show that population imbalanced Fermi-Fermi mixtures 
reduce to strongly interacting (repulsive) Bose-Fermi mixtures in the molecular limit, 
in sharp contrast with homogenous systems where the resulting Bose-Fermi mixtures
are weakly interacting~\cite{pieri,iskin-mixture,iskin-mixture2}.
This result is a direct manifestation of the Pauli exclusion principle in the lattice case, 
since each Bose molecule consists of two fermions, and more than one identical 
fermion on the same lattice site is not allowed for single band systems.
This effect together with the Hartree energy shift lead to a filling dependent
condensate fraction and to sound velocities which
do not approach to zero in the strong attraction limit, in contrast with
the homogeneous case~\cite{iskin-mixture2}.
Lastly, several insulating phases appear in the molecular limit depending 
on the filling fraction and population imbalance.
For instance, we find a molecular Bose-Mott insulator (superfluid) when the molecular filling 
fraction is equal to (less than) one 
for a population balanced gas where the fermion filling fractions are identical. 
When the filling fraction of one type of fermion is one and the filling fraction of the other
is one-half (corresponding to molecular boson and excess fermion filling fractions of one-half),
we also find either a phase-separated state consisting of a Fermi-Pauli insulator (FPI) 
of the excess fermions and a molecular Bose-Mott insulator (BMI) or a Bose-Fermi checkerboard 
(BFC) phase depending on the tunneling anisotropy $\eta$.

The rest of the manuscript is organized as follows.
After introducing the Hamiltonian in Sec.~\ref{sec:hamiltonian}, first
we derive the saddle point self-consistency equations, and then 
analyze the saddle point phase diagrams at zero temperature in Sec.~\ref{sec:saddle}.
In Sec.~\ref{sec:gaussian}, we discuss Gaussian fluctuations at zero temperature
to obtain the low energy collective excitations,
and near the critical temperature to derive the time-dependent Ginzburg-Landau 
(TDGL) equations.
In Sec.~\ref{sec:molecular}, we derive an effective Bose-Fermi action
in the strong attraction limit, and describe the emergence of
insulating BMI, BFC and separated BMI/FPI phases, and in Sec.~\ref{sec:detection}
we suggest some experiments to detect these insulating phases.
A brief summary of our conclusions is given in Sec.~\ref{sec:conclusions}.
Lastly, in App.~\ref{sec:app.a} and~\ref{sec:app.b}, we present 
the elements of the inverse fluctuation matrix, and 
their low frequency and long wavelength expansion coefficients 
at zero and finite temperatures, respectively.

\section{Lattice Hamiltonian}
\label{sec:hamiltonian}

To describe mixtures of fermions loaded into optical lattices, we start with
the following continuum Hamiltonian in real space

\begin{eqnarray}
H_C &=& \sum_\sigma  \int d\mathbf{r} \psi_{\sigma}^\dagger (\mathbf{r}) 
\left[ - \frac {\nabla^2}{ 2m_{\sigma} }  + V_\sigma(\mathbf{r}) - \mu_{\sigma} \right]
\psi_{\sigma} (\mathbf{r}) \nonumber \\
&-& \int d\mathbf{r} \int d\mathbf{r'} \widehat{n}_\uparrow(\mathbf{r}) U(\mathbf{r},\mathbf{r'}) \widehat{n}_\downarrow (\mathbf{r'}),
\label{eqn:hamiltonian.continuum}
\end{eqnarray}
where $\psi_\sigma^\dagger(\mathbf{r})$ [$\psi_\sigma(\mathbf{r})$] field operator 
creates (annihilates) a fermion at position $\mathbf{r}$ with a pseudo-spin $\sigma$,
mass $m_\sigma$ and chemical potential $\mu_\sigma$.
Here,
$
V_\sigma(\mathbf{r})$ is the optical lattice potential,
$
\widehat{n}_\sigma(\mathbf{r}) = \psi_{\sigma}^\dagger (\mathbf{r}) \psi_{\sigma} (\mathbf{r})
$
is the density operator, and
$U(\mathbf{r},\mathbf{r'})$ describes the density-density interaction between fermions.
We allow fermions to be of different species through $m_\sigma$, and
to have different populations controlled by independent $\mu_\sigma$,
where the pseudo-spin $\sigma$ labels the trapped hyperfine states of a given 
species of fermions, or labels different types of fermions in a two-species mixture.
The optical lattice potential has the form 
$
V_\sigma(\mathbf{r}) = V_{0,\sigma} \sum_{j = \{x,y,z\}} \sin^2(\kappa r_j),
$
where $V_{0,\sigma}$ is proportional to the laser intensity, $\kappa = 2\pi/\lambda$ is the 
wavelength of the laser such that $a_c = \lambda/2$ is the size of the lattice spacing,
and $r_j$ corresponds to the $j$th component of $\mathbf{r}$.
This potential describes a cubic optical lattice since the amplitude
and the period of the potential is the same in all three orthogonal directions.
Furthermore, we assume short-range interactions and set 
$U(\mathbf{r},\mathbf{r'}) = U \delta(\mathbf{r}-\mathbf{r'})$, where 
$U > 0$ is the strength of the attractive interactions.

For this Hamiltonian, the single particle eigenfunctions are Bloch wave functions,
leading to a set of Wannier functions that are localized on the individual lattice sites~\cite{wannier}.
Therefore, we expand the creation and annihilation field operators in the 
basis set of these Wannier functions $W_\sigma(\mathbf{r}-\mathbf{r_i})$ of the 
lowest energy states of the optical potential near their minima, such that
$
\psi_\sigma(\mathbf{r}) = \sum_{i} W_\sigma(\mathbf{r}-\mathbf{r_i}) a_{i,\sigma}
$
and
$
\psi_\sigma^\dagger(\mathbf{r}) = \sum_{i} W_\sigma^*(\mathbf{r}-\mathbf{r_i}) a_{i,\sigma}^\dagger,
$
where $a_{i,\sigma}^\dagger$ ($a_{i,\sigma}$) site operator creates (annihilates) a 
fermion at lattice site $i$ with a pseudo-spin $\sigma$.
Here, $\mathbf{r_i}$ is position of the lattice site $i$.
Multi-bands are important when the interaction energy involved in the system is comparable to the
excitation energies to the higher bands, and these effects can be easily incorporated into our theory.
Notice that, for a complete set of Wannier functions such that
$
\sum_i W_\sigma(\mathbf{r}-\mathbf{r_i}) W_\sigma^*(\mathbf{r'}-\mathbf{r_i}) =  \delta(\mathbf{r}-\mathbf{r'}),
$
the field operators
$
[\psi_\sigma(\mathbf{r}), \psi_{\sigma'}^\dagger(\mathbf{r'})] = \delta_{\sigma,\sigma'} \delta(\mathbf{r}-\mathbf{r'})
$
as well as site operators 
$
[a_{i,\sigma}, a_{j,\sigma'}^\dagger] = \delta_{\sigma,\sigma'} \delta_{i,j}
$
obey the Fermi anti-commutation rules, where $\delta(\mathbf{r})$ is the delta function
and $\delta_{i,j}$ is the Kronecker-delta.

This expansion reduces the continuum Hamiltonian given in Eq.~(\ref{eqn:hamiltonian.continuum})
to a site Hamiltonian
\begin{eqnarray}
H_S &=& - \sum_{i,j,\sigma}  t_\sigma^{i,j} \widehat{n}_{i,j;\sigma}
- \sum_{i,\sigma}  \mu_\sigma \widehat{n}_{i,i;\sigma} \nonumber \\
&-& \sum_{i,j,k,l} \widehat{n}_{i,j;\uparrow} U_{i,j;k,l} \widehat{n}_{k,l;\downarrow},
\label{eqn:hamiltonian.site}
\end{eqnarray}
where 
$
t_\sigma^{i,j} = -\int d\mathbf{r} W_\sigma^*(\mathbf{r}-\mathbf{r_i})
[ - \nabla^2 /(2m_{\sigma})  + V_\sigma(\mathbf{r}) ]
W_\sigma(\mathbf{r}-\mathbf{r_j})
$
is the tunneling amplitude between sites $i$ and $j$,
$
U_{i,j;k,l} = U \int d\mathbf{r}  W_\uparrow^*(\mathbf{r}-\mathbf{r_i}) W_\uparrow(\mathbf{r}-\mathbf{r_j})
W_\downarrow^*(\mathbf{r}-\mathbf{r_k}) W_\downarrow(\mathbf{r}-\mathbf{r_l})
$
is the elements of strength of attractive density-density interactions 
between sites $\{i,j\}$ and $\{k,l\}$, and
$
\widehat{n}_{i,j;\sigma} = a_{i,\sigma}^\dagger a_{j,\sigma}.
$
In this manuscript, we allow tunneling and interactions up to the nearest-neighbors, 
and choose
$
t_\sigma^{i,j} = t_\sigma \delta_{i, j \pm 1}
$
and
$
U_{i,j;k,l} = g \delta_{i,j} \delta_{j,k} \delta_{k,l} + h \delta_{i,j} \delta_{j,k \pm 1} \delta_{k,l},
$
respectively.
Here, $g > 0$ and $h > 0$ are strengths of the on-site and nearest-neighbor 
interactions, respectively. 
This reduces the general site Hamiltonian given in Eq.~(\ref{eqn:hamiltonian.site})
to a nearest-neighbor site Hamiltonian
\begin{eqnarray}
H_S^{NN} &=& - \sum_{\langle i,j \rangle,\sigma}  t_\sigma \widehat{n}_{i,j;\sigma}
- \sum_{i,\sigma}  \mu_\sigma \widehat{n}_{i,i;\sigma} \nonumber \\
&-& g \sum_{i} \widehat{n}_{i,i;\uparrow} \widehat{n}_{i,i;\downarrow}
- h \sum_{\langle i,j \rangle} \widehat{n}_{i,i;\uparrow} \widehat{n}_{j,j;\downarrow},
\label{eqn:hamiltonian.siteNN}
\end{eqnarray}
where $\langle i, j \rangle$ restricts sums to nearest-neighbors only such that $j = i \pm 1$.
This is the nearest-neighbor Fermi-Hubbard Hamiltonian for a simple cubic lattice.

Finally, we Fourier transform the site operators to the momentum space ones, such that
$
a_{i,\sigma} = (1/\sqrt{M}) \sum_\mathbf{k} a_{\mathbf{k}, \sigma} e^{-i \mathbf{k} \cdot \mathbf{r_i}}
$
and
$
a_{i,\sigma}^\dagger = (1/\sqrt{M}) \sum_\mathbf{k} a_{\mathbf{k}, \sigma}^\dagger e^{i \mathbf{k} \cdot \mathbf{r_i}},
$
where $M$ is the number of lattice sites, and $a_{\mathbf{k}, \sigma}^\dagger$ ($a_{\mathbf{k}, \sigma}$) 
operator creates (annihilates) a fermion at momentum $\mathbf{k}$ with a pseudo-spin $\sigma$. 
This leads to the general momentum space Hamiltonian
\begin{eqnarray}
H_M &=& \sum_{\mathbf{k},\sigma}\xi_{\mathbf{k},\sigma} \widehat{n}_{\mathbf{k},\mathbf{k}; \sigma}  \nonumber \\
&-& \sum_{\mathbf{k_1}+\mathbf{k_3} = \mathbf{k_2}+\mathbf{k_4}}
\widehat{n}_{\mathbf{k_1},\mathbf{k_2}; \uparrow} U(\mathbf{k_1}-\mathbf{k_2}) \widehat{n}_{\mathbf{k_3},\mathbf{k_4}; \downarrow},
\label{eqn:hamiltonian.momentum}
\end{eqnarray}
where
$
\xi_{\mathbf{k},\sigma}= \epsilon_{\mathbf{k},\sigma} - \mu_\sigma
$ 
describes the nearest-neighbor tight-binding dispersion
$
\epsilon_{\mathbf{k},\sigma} = -2t_\sigma \sum_{j = \{x,y,z\}} \cos (k_j a_c)
$
where $k_j$ is the $j$th component of $\mathbf{k}$, and $a_c$ is the lattice spacing.
Here, 
$
\widehat{n}_{\mathbf{k},\mathbf{k'}; \sigma} = a_{\mathbf{k}, \sigma}^\dagger a_{\mathbf{k'}, \sigma},
$
and we used
$
\sum_{\langle i, j \rangle} e^{i \mathbf{k} \cdot (\mathbf{r_i} - \mathbf{r_j})} = 2\sum_{j=\{x,y,z\}} \cos (k_j a_c).
$
Therefore, in order to achieve the momentum space Hamiltonian given in Eq.~(\ref{eqn:hamiltonian.momentum}) 
from the real space Hamiltonian given in Eq.~(\ref{eqn:hamiltonian.continuum}), 
we performed
$
\psi_\sigma(\mathbf{r}) = (1/\sqrt{M}) \sum_{i,\mathbf{k}} a_{\mathbf{k},\sigma} 
W_\sigma(\mathbf{r}-\mathbf{r_i}) e^{-i \mathbf{k} \cdot \mathbf{r_i}}
$
and
$
\psi_\sigma^\dagger(\mathbf{r}) = (1/\sqrt{M}) \sum_{i,\mathbf{k}} a_{\mathbf{k},\sigma}^\dagger 
W_\sigma^*(\mathbf{r}-\mathbf{r_i}) e^{i \mathbf{k} \cdot \mathbf{r_i}},
$
which corresponds to the total transformation from the real space 
field operators to the momentum space ones.

The second term in Eq.~(\ref{eqn:hamiltonian.momentum})
$
U(\mathbf{k}-\mathbf{k'}) = (1/M) \sum_{i,j} U_{i,i;j,j} e^{(\mathbf{k}-\mathbf{k'}) \cdot (\mathbf{r_i}-\mathbf{r_j})}
$ 
is the Fourier transform of the interaction term and is given by
$
U(\mathbf{k}-\mathbf{k'}) = g + 2h \sum_{j = \{x,y,z\}} \cos[(k_j - k_j')a_c].
$
In momentum space, the simplification comes from the separability of the
interaction potential, such that
$
U(\mathbf{k}-\mathbf{k'}) = g \Gamma_s (\mathbf{k}) \Gamma_s (\mathbf{k'}) +
h \sum_{j = \{e,d,p\}} \Gamma_j (\mathbf{k}) \Gamma_j (\mathbf{k'})
$
where the sum over $c$ corresponds to different pairing symmetries. 
For a three-dimensional lattice, we obtain the following components:
$
\Gamma_s(\mathbf{k}) = 1
$
for the s-wave symmetry;
$
\Gamma_e(\mathbf{k}) = [\cos(k_x a_c) + \cos(k_y a_c) + \cos(k_z a_c)]/\sqrt{2}
$
for the extended s-wave symmetry;
$
\Gamma_{d,1}(\mathbf{k}) = [\cos(k_x a_c) + \cos(k_y a_c) - \cos(k_z a_c)]/\sqrt{2},
$
$
\Gamma_{d,2}(\mathbf{k}) = [\cos(k_x a_c) - \cos(k_y a_c) + \cos(k_z a_c)]/\sqrt{2}
$
and
$
\Gamma_{d,3}(\mathbf{k}) = [\cos(k_x a_c) - \cos(k_y a_c) - \cos(k_z a_c)]/\sqrt{2}
$
for the d-wave symmetry; and
$
\Gamma_{p,1}(\mathbf{k}) = \sqrt{2} \sin(k_x a_c),
$
$
\Gamma_{p,2}(\mathbf{k}) = \sqrt{2} \sin(k_y a_c)
$
and
$
\Gamma_{p,3}(\mathbf{k}) = \sqrt{2} \sin(k_z a_c)
$
for the p-wave symmetry. Notice that, symmetrized and anti-symmetrized 
combinations of cosines and sines help to exploit the symmetry of the lattice.
In this manuscript, we set nearest-neighbor attraction to zero ($h = 0$),
and consider only the s-wave on-site attractions ($g \ne 0$).

Thus, to describe mixtures of fermions loaded into optical lattices, we use the s-wave 
single-band Hamiltonian
\begin{eqnarray}
\label{eqn:hamiltonian}
H = \sum_{\mathbf{k},\sigma}\xi_{\mathbf{k},\sigma} a_{\mathbf{k}, \sigma}^\dagger a_{\mathbf{k}, \sigma} 
- g \sum_{\mathbf{k},\mathbf{k',\mathbf{q}}} b_{\mathbf{k},\mathbf{q}}^\dagger b_{\mathbf{k'},\mathbf{q}}, 
\end{eqnarray}
with an on-site attractive interaction $g > 0$.  
Here, $a_{\mathbf{k}, \sigma}^\dagger$ is the fermion creation and
$b_{\mathbf{k},\mathbf{q}}^\dagger = a_{\mathbf{k}+\mathbf{q}/2,\uparrow}^\dagger 
a_{-\mathbf{k}+\mathbf{q}/2,\downarrow}^\dagger \Gamma_s^* (\mathbf{k})$ is the pair creation operator.
In addition,
$
\xi_{\mathbf{k},\sigma}= \epsilon_{\mathbf{k},\sigma} - \widetilde{\mu}_\sigma
$ 
describes the nearest-neighbor tight-binding dispersion,
$
\epsilon_{\mathbf{k},\sigma} = 2t_\sigma \theta_{\mathbf{k}}
$ 
with
\begin{equation}
\theta_{\mathbf{k}} = \sum_{j=\{x,y,z\}} [1 - \cos (k_j a_c)],
\label{eqn:dispersion}
\end{equation}
where $\widetilde{\mu}_\sigma = \mu_\sigma - V_{H,\sigma}$ and $V_{H,\sigma}$ is a possible Hartree energy shift.
Notice that, we allow fermions to be of different species through $t_\sigma$, and
to have different populations controlled by independent $\widetilde{\mu}_\sigma$.

Furthermore, the momentum space sums in Eq.~(\ref{eqn:hamiltonian}) and throughout this 
manuscript are evaluated as follows. For large and translationally invariant systems
considered here, we can use a continuum approximation to sum over the discrete $\mathbf{k}$ levels, 
and for a three dimensional lattice write
\begin{eqnarray}
\sum_\mathbf{k} F(k_x, k_y, k_z)
&\equiv& \int_{-\pi/a_c}^{\pi/a_c} \frac{dk_x dk_y dk_z}{(2\pi/L)^3} F(k_x, k_y, k_z), \nonumber \\
&=& M \int_{-\pi}^{\pi} \frac{d\widetilde{k}_x d\widetilde{k}_y d\widetilde{k}_z}{(2\pi)^3} 
F(\widetilde{k}_x, \widetilde{k}_y, \widetilde{k}_z),
\end{eqnarray}
where $F(k_x, k_y, k_z)$ is a generic function of $k_x$, $k_y$ and $k_z$, 
$L = M a_c$ is the size of the lattice, and $\widetilde{k_i} = k_i a_c$ is dimensionless. 
This continuum approximation is valid for large systems only where $M \gg 1$.
Notice also that, $a_c$ provides a natural cutoff to the $\mathbf{k}$ space integrations 
in the lattice case.

Unlike recent works on fermion pairing in optical lattices which were restricted to 
the BCS limit~\cite{liu-lattice,torma-lattice},
we discuss next the evolution of superfluidity from the BCS to the BEC regime~\cite{iskin-prb}
and the emergence of insulating phases~\cite{iskin-lattice}.
We ignore multi-band effects since a single-band Hamiltonian 
may be sufficient to describe the evolution from BCS to BEC physics
in optical lattices~\cite{stoof-2006}. 
Multi-bands are important when the fermion filling fraction
is higher than one and/or the optical lattice is not in the tight-binding regime,
but these effects can be easily incorporated into our theory.

\section{Saddle Point Approximation}
\label{sec:saddle}

In this manuscript, we use the functional integral formalism described in Ref.~\cite{carlos, jan, iskin-mixture, iskin-mixture2}.
The general method follows the same prescription as for the homogeneous case~\cite{iskin-mixture2}, and we
do not repeat the same analysis for the lattice case discussed here. The main differences between the 
momentum space expressions for the homogeneous and the lattice case
come from the periodic dispersion relation 
for the lattice system, as given in Eq.~(\ref{eqn:dispersion}). 

\subsection{Saddle Point Self-Consistency Equations}
\label{sec:self-consistency}

For the Hamiltonian given in Eq.~(\ref{eqn:hamiltonian}), 
the saddle point order parameter equation parallels that of the homogeneous 
system~\cite{iskin-mixture2}, and leads to
\begin{equation}
\frac{1}{g} = \frac{1}{M}\sum_{\mathbf{k}} \frac{1 - f(E_{\mathbf{k},1}) - f(E_{\mathbf{k},2})}
{2E_{\mathbf{k},+}}  |\Gamma_s (\mathbf{k})|^2,
\label{eqn:op}
\end{equation}
where $M$ is the number of lattice sites,
$
f(x) = 1/[\exp(x/T) + 1]
$
is the Fermi function,
\begin{equation}
E_{\mathbf{k},s} = (\xi_{\mathbf{k},+}^2 + |\Delta_{\mathbf{k}}|^2)^{1/2} + \gamma_s \xi_{\mathbf{k},-}
\end{equation}
is the quasiparticle energy when $\gamma_1 = 1$ or
the negative of the quasihole energy when $\gamma_2 = -1$, and
$
E_{\mathbf{k},\pm} = (E_{\mathbf{k},1} \pm E_{\mathbf{k},2})/2.
$
Here, $\Delta_{\mathbf{k}} = \Delta_0 \Gamma_s (\mathbf{k})$ is the order parameter and
$
\xi_{\mathbf{k},\pm} = \epsilon_{\mathbf{k},\pm} - \widetilde{\mu}_\pm,
$
where 
$
\epsilon_{\mathbf{k},\pm} = 2t_\pm \theta_{\mathbf{k}}
$
with $t_\pm = (t_\uparrow \pm t_\downarrow)/2$ and 
$
\widetilde{\mu}_\pm = (\widetilde{\mu}_\uparrow \pm \widetilde{\mu}_\downarrow)/2.
$
Notice that the symmetry between quasiparticles and quasiholes 
is broken when $\xi_{\mathbf{k},-} \ne 0$. 
The order parameter equation has to be solved self-consistently 
and leads to number equations
\begin{eqnarray}
N_{\uparrow} &=& \sum_{\mathbf{k}} \left[ |u_{\mathbf{k}}|^2 f(E_{\mathbf{k},1})+ |v_{\mathbf{k}}|^2 f(-E_{\mathbf{k},2}) \right],
\label{eqn:numberup} \\
N_{\downarrow} &=& \sum_{\mathbf{k}} \left[ |u_{\mathbf{k}}|^2 f(E_{\mathbf{k},2})+ |v_{\mathbf{k}}|^2 f(-E_{\mathbf{k},1}) \right],
\label{eqn:numberdo}
\end{eqnarray}
which are derived the same way as in the homogeneous case~\cite{iskin-mixture2}.
Here,  
$
|u_{\mathbf{k}}|^2 = (1 + \xi_{\mathbf{k}, +}/E_{\mathbf{k}, +})/2
$
and
$
|v_{\mathbf{k}}|^2 = (1 - \xi_{\mathbf{k}, +}/E_{\mathbf{k}, +})/2.
$
The number of $\sigma$-type fermions per lattice site is given by 
\begin{equation}
0 \le n_\sigma = \frac{N_\sigma}{M} \le 1.
\end{equation}
Thus, when $n_\uparrow \ne n_\downarrow$, we need to solve all
three self-consistency equations, since population imbalance is achieved
when either $E_{\mathbf{k},1}$ or $E_{\mathbf{k},2}$ is negative 
in some regions of momentum space, as discussed next.

\subsection{Saddle Point Phase Diagrams \\ at Zero Temperature}
\label{sec:phase.diagrams}

\begin{figure} [htb]
\centerline{\scalebox{0.6} {\includegraphics{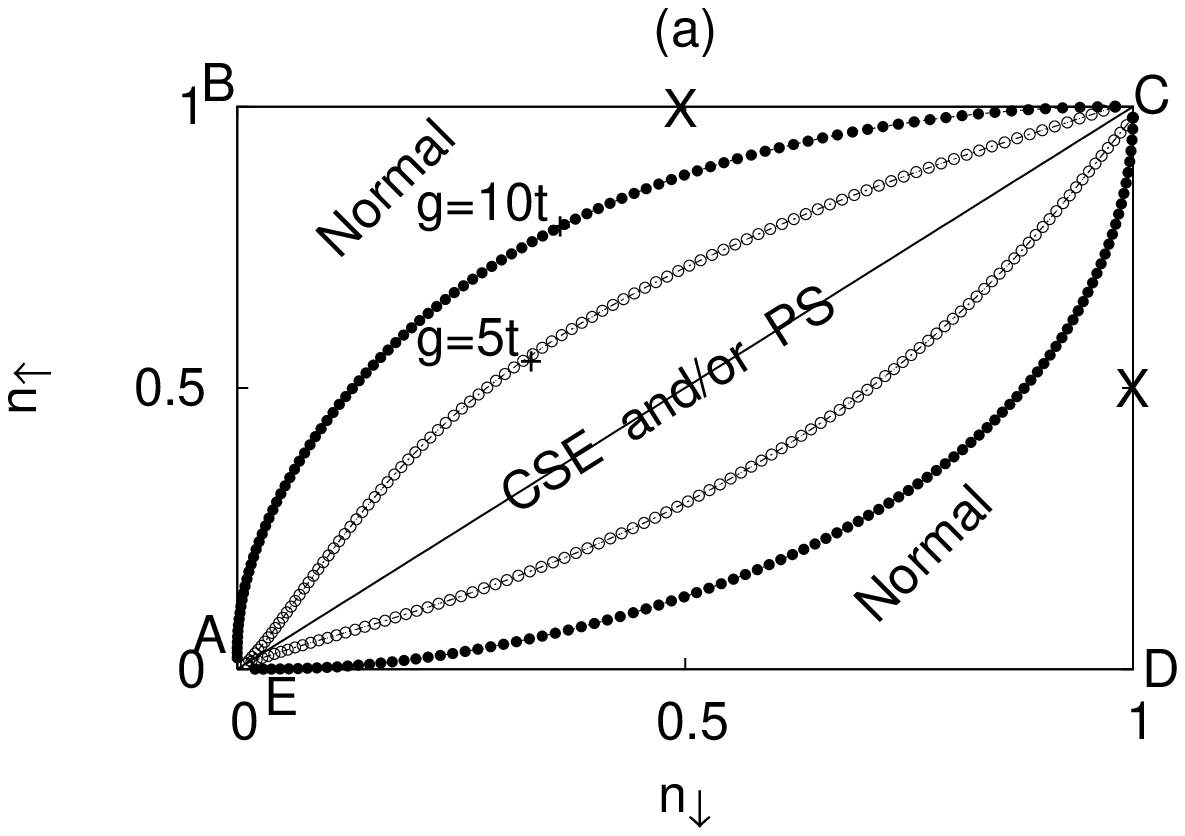}}}
\centerline{\scalebox{0.6} {\includegraphics{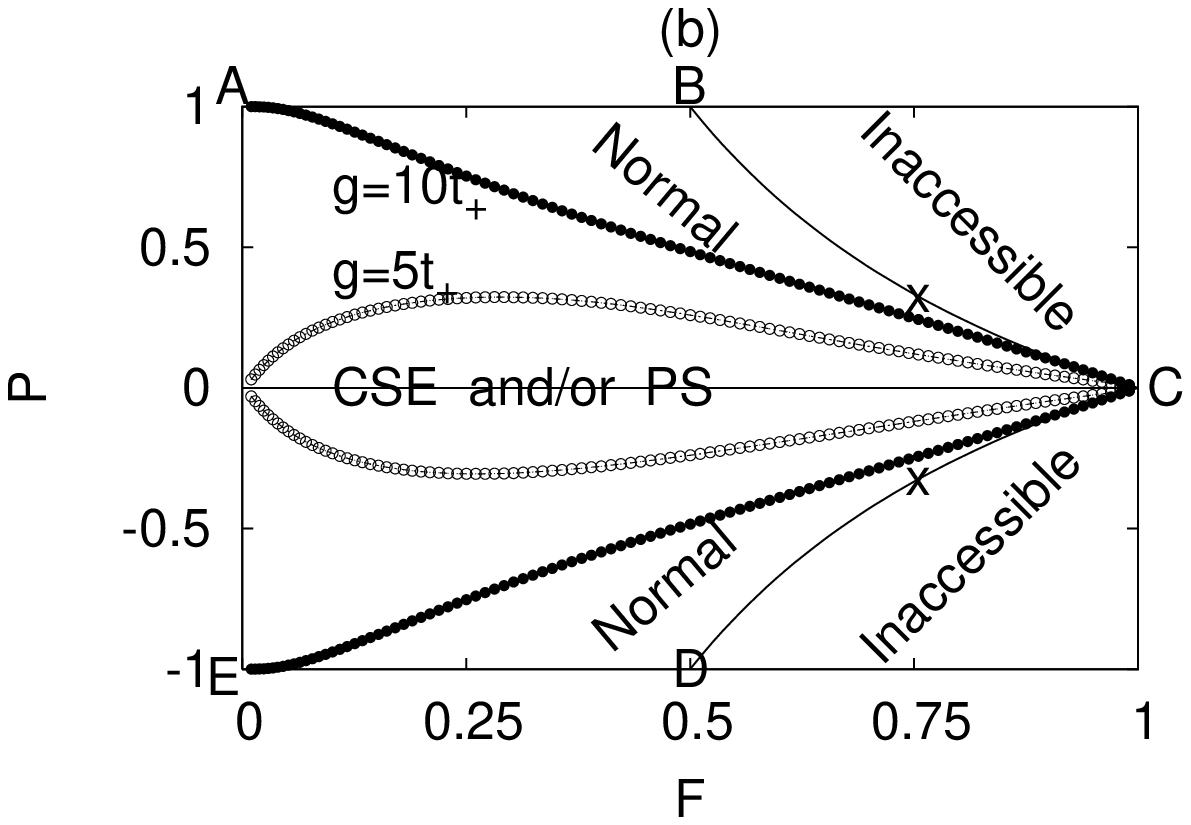}}}
\caption{\label{fig:lattice.tr.1}
Phase diagrams for a one-species ($^6$Li or $^{40}$K) mixture of two-hyperfine states with $\eta = 1$:
(a) $n_\uparrow$ versus $n_\downarrow$, and
(b) $P$ versus $F$,
for $g = 5t_+$ and $g = 10t_+$. The normal regions (outside the `football' boundaries) 
and coexistence of superfluidity with excess fermions (CSE) and/or 
phase separation (PS) (inside the `football' boundaries) are indicated. 
The CSE/PS (normal) region expands (shrinks) with increasing fermion attraction. 
}
\end{figure}

\begin{figure} [htb]
\centerline{\scalebox{0.6} {\includegraphics{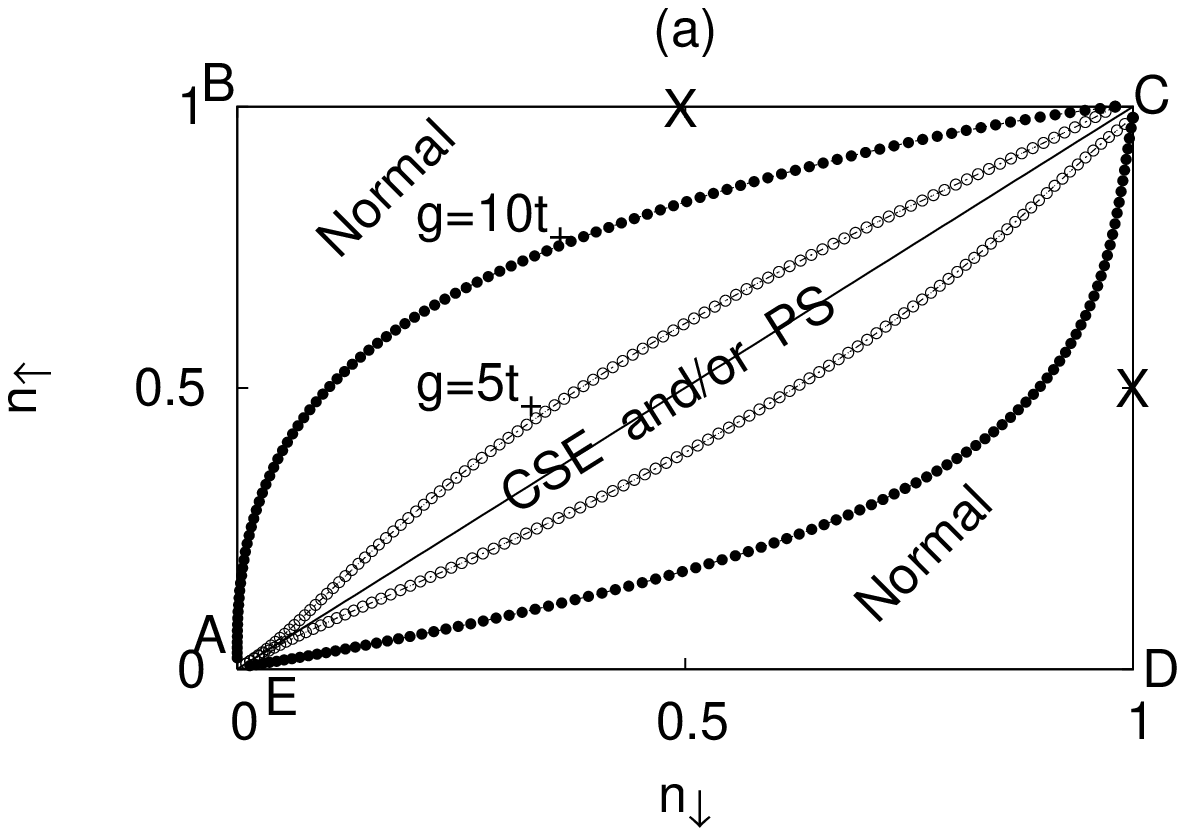}}}
\centerline{\scalebox{0.6} {\includegraphics{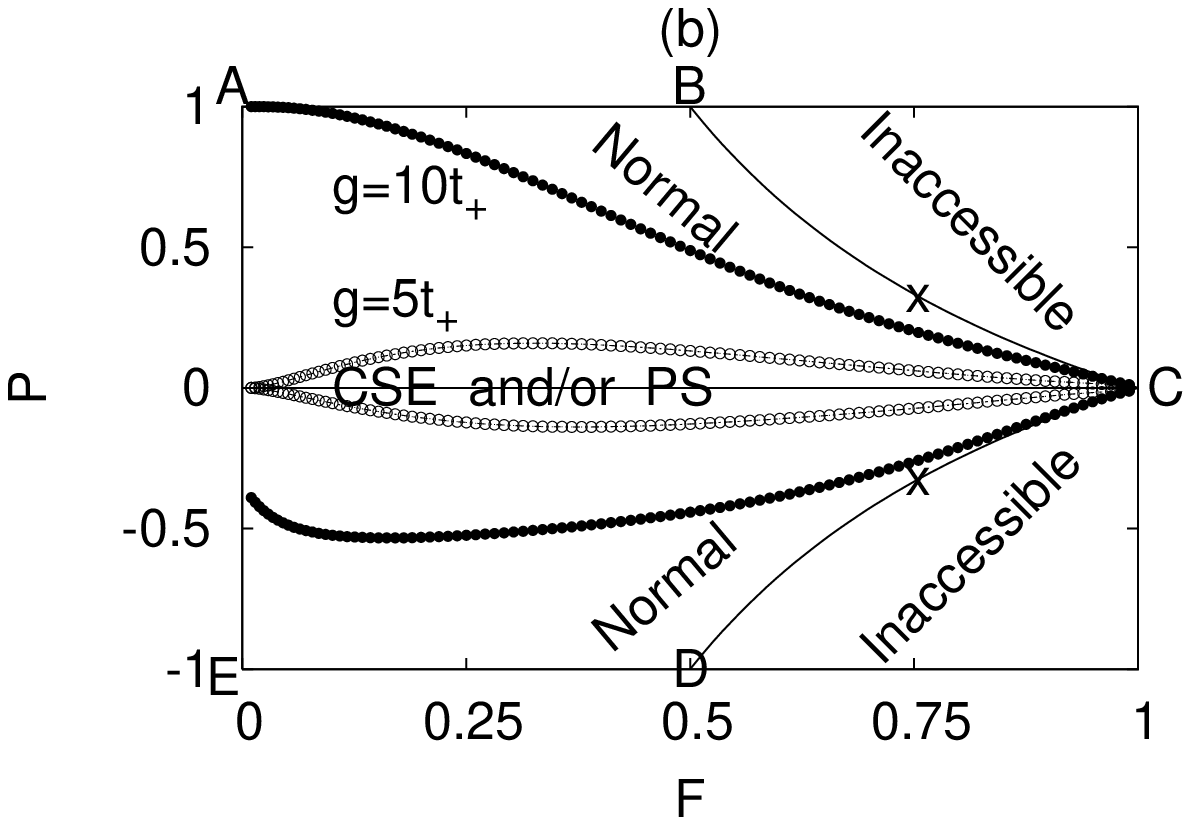}}}
\caption{\label{fig:lattice.tr.6.67}
Phase diagrams for a two-species ($^6$Li and $^{40}$K) mixture of two-hyperfine states with $\eta = 0.15$:
(a) $n_\uparrow$ versus $n_\downarrow$, and
(b) $P$ versus $F$,
for $g = 5t_+$ and $g = 10t_+$. The normal regions (outside the `football' boundaries) 
and coexistence of superfluidity with excess fermions (CSE) and/or 
phase separation (PS) (inside the `football' boundaries) are indicated. 
The CSE/PS (normal) region expands (shrinks) with increasing fermion attraction. 
}
\end{figure}

To obtain ground state phase diagrams, 
we solve Eqs.~(\ref{eqn:op}),~(\ref{eqn:numberup}) and ~(\ref{eqn:numberdo})
as a function of interaction strength $g$, population imbalance 
and total filling fraction
\begin{eqnarray}
-1 \le P &=& \frac{n_\uparrow - n_\downarrow}{n_\uparrow + n_\downarrow} \le 1, \\
0 \le F &=& \frac{n_\uparrow + n_\downarrow}{2} \le 1,
\end{eqnarray}
respectively~\cite{iskin-lattice}, and consider two sets of tunneling ratios $\eta = t_\downarrow / t_\uparrow$.
The case of $\eta = 1$ is shown in Fig.~\ref{fig:lattice.tr.1}, and 
the case of $\eta = 0.15$ is shown in Fig.~\ref{fig:lattice.tr.6.67}.
While $\eta = 1$ corresponds to a one-species (two-hyperfine-state) mixture such
as $^6$Li or $^{40}$K, $\eta = 0.15$ corresponds to a two-species mixture (one-hyperfine-state of each
type of atom) such as $^6$Li and $^{40}$K; $^6$Li and $^{87}$Sr; or $^{40}$K and $^{87}$Sr.

Generally, lines AB ($0 < n_\uparrow < 1$; $n_{\downarrow} = 0$) and 
ED ($n_\uparrow = 0$; $ 0 < n_{\downarrow} < 1$) 
in Figs.~\ref{fig:lattice.tr.1} and~\ref{fig:lattice.tr.6.67}, 
correspond to normal $\sigma$-type Fermi metal for all interactions, 
while points B $(n_\uparrow = 1, n_\downarrow = 0)$ and D $(n_\uparrow = 0, n_\downarrow = 1)$
correspond to a Fermi-Pauli (band) insulator since there is only one type of
fermion in a fully occupied band. Thus, the only option for additional 
fermions $(\uparrow$ in case B and $\downarrow$ in case D) 
is to fill higher energy bands if the optical potential supports it, 
otherwise the extra fermions are not trapped. For the case where no additional bands are occupied, 
we label the corresponding phase diagram regions as `Inaccessible' 
in Figs.~\ref{fig:lattice.tr.1}(b) and~\ref{fig:lattice.tr.6.67}(b), 
since either $n_\uparrow > 1$ or $n_\downarrow > 1$ in these regions.
This phase boundary between the `Normal' and the `Inaccessible' phase is given by 
$P = 1/F - 1$ for $P > 0$, and $P = 1 - 1/F$ for $P < 0$.

The population balanced line AC ends at the special point C, 
where $n_\uparrow = n _\downarrow = 1$. This point is a Fermi-Pauli (band) insulator
for weak attraction since both fermion bands are fully occupied.
Furthermore, for very weak attraction, lines 
BC ($n_\uparrow = 1, 0 < n_\downarrow < 1$) and CD 
($0 < n_\uparrow < 1, n_\downarrow = 1$) correspond essentially 
to a fully polarized ferromagnetic metal (or half-metal), 
where only the type of fermions with filling fraction less than one can move around.

In the phase diagrams shown in Figs.~\ref{fig:lattice.tr.1} and~\ref{fig:lattice.tr.6.67}, 
we indicate the regions of normal (N) phase where $|\Delta_0| = 0$, and group
together the regions of coexistence of superfluidity and excess fermions (CSE) 
and/or phase separation (PS), where $|\Delta_0| \ne 0$.
When $F \ll 1$, the phase diagrams are similar to the homogenous case~\cite{pao, sheehy, iskin-mixture, iskin-mixture2}, 
and the $P$ versus $F$ phase diagram is symmetric for equal tunnelings as shown 
in Fig.~\ref{fig:lattice.tr.1}(b), and is asymmetric for unequal tunnelings having 
a smaller normal region when the lighter band mass fermions are in excess 
as shown in Fig.~\ref{fig:lattice.tr.6.67}(b).
Here, we do not discuss separately the CSE and PS regions since they have already been 
discussed in homogeneous and harmonically trapped systems~\cite{pao, sheehy, iskin-mixture, iskin-mixture2}, 
and experimentally observed~\cite{mit, rice, mit-2, rice-2}, but we make two remarks.

First, the lattice phase diagram is also different from the homogenous systems' in relation
to the topological quantum phase transitions discussed in Ref.~\cite{iskin-mixture, iskin-mixture2}.
These phases are characterized by the number of zero-energy surfaces
of $E_{\mathbf{k},1}$ and $E_{\mathbf{k},2}$ such that
(I) $E_{\mathbf{k},1(2)}$ has no zeros and $E_{\mathbf{k},2(1)}$ has only one,
(II) $E_{\mathbf{k},1(2)}$ has no zeros and $E_{\mathbf{k},2(1)}$ has two zeros, and
(III) $E_{\mathbf{k},1(2)}$ and $E_{\mathbf{k},2(1)}$ has no zeros 
and are always positive corresponding to the $P = 0$ limit.
The topological phases characterized by the number (I and II) of simply connected 
zero-energy surfaces of $E_{\mathbf{k},s}$ may lie in the stable region of CSE,
unlike in the homogeneous case where the topological phase II 
always lies in the phase separated region for all parameter space~\cite{iskin-mixture, iskin-mixture2}. 

To understand the topology of the uniform superfluid phase, we define 
$
B_\pm = \widetilde{\mu}_\uparrow/(4t_\uparrow) \pm \widetilde{\mu}_\downarrow/(4t_\downarrow).
$
For the s-wave symmetry considered, the zeros of $E_{\mathbf{k},s}$ occur at real momenta
$
\theta_{\mathbf{k}}^\pm = B_+ \pm [B_-^2 - |\Delta_0|^2/(4t_\uparrow t_\downarrow)]^{1/2}
$
provided that 
$
|\Delta_0|^2 < 4 |B_-|^2 t_\uparrow t_\downarrow
$
for $B_+ \ge 0$, and 
$
|\Delta_0|^2 < - \widetilde{\mu}_\uparrow \widetilde{\mu}_\downarrow
$
for $B_+ < 0$.
The transition from phase II to I occurs when $\theta_\mathbf{k}^- \to 0$,
indicating a change in topology in the lowest quasiparticle band.
The topology is of phase I if $B_+ \le 0$, and phase II is possible only when $B_+ > 0$, 
and therefore phase II (I) always appears in the BCS (BEC) side of the phase diagram.
In the particular case when $t_\uparrow = t_\downarrow$, 
the topology is of phase I if $\widetilde{\mu}_+ \le 0$,
and phase II is possible only when $\widetilde{\mu}_+ > 0$.

Notice that, $\widetilde{\mu}_+$ goes below the bottom of particle (hole) band as $g$ increases,
since the Cooper pairs are formed from particles (holes) when $F < 0.5$ ($F > 0.5$)~\cite{iskin-prb}.
This leads to a large and negative (positive) $\widetilde{\mu}_+$ for particle (hole) 
pairs in the BEC limit, thus leading to a uniform superfluid with topological phase I (II).
Therefore, the topology of the entire superfluid phase is expected to be of phase II for $F \ge 0.5$.
However, when $t_\uparrow = t_\downarrow$, the boundary between phase II and phase 
I lies around $g \approx 13t_+$ for $F = 0.1$ and $g \sim 30t_+$ for $F = 0.3$.

Similar topological phase transitions in nonzero angular momentum superfluids have 
been discussed previously in the literature in connection with $^3$He p-wave 
phases~\cite{leggett, volovik-book} and cuprate d-wave superconductors~\cite{duncan}, and more
recently a connection to ultracold fermions which exhibit p-wave Feshbach resonances was
made~\cite{volovik-pwave, botelho-pwave, gurarie-pwave, skyip-pwave}.
However, we would like to emphasize that the topological transition discussed here is unique, 
since it involves a stable s-wave superfluid, and may be 
observed for the first time in future experiments.

Second, the phase diagram characterized by normal, superfluid (CSE or PS), and insulating 
regions may be explored experimentally by tuning the ratio $g/t_+$, total filling fraction $F$, 
and population imbalance $P$ as done in harmonic traps~\cite{mit, rice, mit-2, rice-2}.
We would like to emphasize that our saddle point results provide a non-perturbative 
semi-quantitative description of the normal/superfluid phase boundary for all couplings 
and tells us that the system is either superfluid $(\vert \Delta_0 \vert \ne 0)$ 
or normal $(\vert \Delta_0 \vert = 0 )$, but fails to describe the insulating phases. 
Thus, first, we analyze Gaussian fluctuations at zero and finite temperatures, 
and then show that insulating phases emerge from fluctuation effects 
beyond the saddle point approximation.

\section{Gaussian fluctuations}
\label{sec:gaussian}

In this section, we follow closely the formalism developed to deal with fluctuations
for homogeneous superfluids~\cite{carlos, jan, iskin-mixture, iskin-mixture2}.
We determine the collective modes for Fermi superfluids in optical lattices at 
zero temperature and the efffective equation of motion near the critical temperature. 

\subsection{Gaussian Fluctuations \\ at Zero Temperature}
\label{sec:gaussian.zero}

Next, we analyze the zero temperature Gaussian fluctuations in the BCS and BEC limits for $P = 0$,
from which we extract the low frequency and long wavelength collective excitations.
For the s-wave symmetry considered, the collective excitation spectrum 
is determined from the poles of the fluctuation matrix in 
much the same way as in homogenous systems~\cite{iskin-mixture2}. Thus, following 
Ref.~\cite{iskin-mixture2}, we obtain the condition
\begin{equation}
\det \left( \begin{array}{cc} A + C|\mathbf{q}|^2 - Dw^2 & iBw 
\\ -iBw & Q|\mathbf{q}|^2 - Rw^2 \end{array}\right) = 0,
\label{eqn:coll}
\end{equation}
where the expansion coefficients $A, B, C, D, Q$ and $R$ are given in App.~\ref{sec:app.a}.
There are amplitude and phase branches for the collective excitations, 
but we focus only on the lowest energy Goldstone phase mode with dispersion
$
w(\mathbf{q}) = v |\mathbf{q}|,
$
where 
\begin{equation}
v = \left(\frac{A Q}{AR + B^2}\right)^{1/2}
\end{equation}
is the speed of sound. First, we discuss analytically the sound velocity in the 
weak and strong attraction limits, and then calculate numerically the
evolution in between these limits.

In the weak attraction (BCS) limit when $0 < \widetilde{\mu}_+ < 4{\cal D}t_+$,
the coefficient that couples phase and amplitude fields vanish ($B = 0$), 
and the phase and amplitude modes are decoupled.
We also obtain
$
A = {\cal N}(0) ,
$
$
C = Q/3 = {\cal N}(0) v_{F}^2/ (12 {\cal D} |\Delta_0|^2),
$
and
$
D = R/3 = {\cal N}(0) / (12|\Delta_0|^2),
$
where ${\cal D}$ is the number of dimensions.
Here, 
$
{\cal N}(x) = \sum_{\mathbf{k}} \delta(\xi_{\mathbf{k},+} - x)
$
is the density of states and
$
v_{F}^2 = {\cal N}_r(0)/{\cal N}(0)
$ 
is the Fermi velocity, where $\delta(x)$ is the delta function and
$
{\cal N}_r(x) = \sum_{\mathbf{k}, i} 
[(\nabla_i \xi_{\mathbf{k},\uparrow})^2 - (\nabla_i \xi_{\mathbf{k},\downarrow})^2] 
\delta(\xi_{\mathbf{k},+} - x) 
$
is an effective `kinetic' density of states.
In addition, we find
$
|\Delta_0| = 2[(2{\cal D}t_+)^2 - (\widetilde{\mu}_+ - 2{\cal D}t_+)^2]^{1/2} \exp[-1/(g{\cal N}(0))]
$
for the order parameter and
$
\widetilde{\mu}_+ = F/{\cal N}(0)
$
for the chemical potential.

The calculation given above leads to
$
v^2 = v_F^2 / {\cal D},
$
which reduces to the Anderson-Bogoliubov relation when $t_\uparrow = t_\downarrow$.
In this limit, the sound velocity can also be written as
$
v^2 = v_{F,\uparrow} v_{F,\downarrow} / {\cal D}
$
where 
$
v_{F,\sigma}^2 = {\cal N}_{r,\sigma}(0)/{\cal N}_\sigma(0).
$
Here, 
$
{\cal N}_\sigma(x) = \sum_{\mathbf{k}} \delta(\xi_{\mathbf{k},\sigma} - x)
$
is the density of states, and
$
{\cal N}_{r,\sigma}(x) = \sum_{\mathbf{k}, i} 
(\nabla_i \xi_{\mathbf{k},\sigma})^2 \delta(\xi_{\mathbf{k},\sigma} - x).
$

In the strong attraction (BEC) limit when $\widetilde{\mu}_+ < 0$ for $F < 0.5$ 
and $\widetilde{\mu}_+ > 4{\cal D}t_+$ for $F > 0.5$,
the coefficient $B \ne 0$ indicates that the amplitude and phase fields are coupled.
We obtain
$
A = 4F (1-F) / g,
$
$
B = (1-2F) / g^2,
$
$
D = (1-2F)^2 / g^3,
$
$
Q = 2a_c^2 t_\uparrow t_\downarrow / g^3
$
and
$
R = 1 / g^3,
$
where we used 
$
|\Delta_0| = g [F (1-F)]^{1/2}
$
for the order parameter and
$
\widetilde{\mu}_+ = -(g/2 + 4{\cal D} t_\uparrow t_\downarrow/g) (1-2F) - 2{\cal D}t_+
$
for the average chemical potential.
Notice that, $\mu_+ = \epsilon_b/2 - 2{\cal D}t_+ - 4{\cal D}t_\uparrow t_\downarrow(1-2F)/g$, where $\epsilon_b = - g$ is the
binding energy defined by
\begin{equation}
\frac{1}{g} = \sum_{\mathbf{k}}\frac{|\Gamma_s (\mathbf{k})|^2}{2\epsilon_{\mathbf{k},+} - \epsilon_b},
\end{equation}
and $V_{H} = gF$ is the Hartree energy.
Thus, we obtain
$
v^2 = 8a_c^2 t_\uparrow t_\downarrow F (1-F) = v_\uparrow v_\downarrow
$
for the sound velocity, where 
$
v_\sigma = 2a_c t_\sigma \sqrt{2F (1-F)}.
$
Notice that, $v$ has a $\sqrt{F}$ dependence in the dilute limit 
which is consistent with the homogenous result where the sound velocity
depends on the square-root of the density. 
However, $v$ saturates to a finite value as $g$ increases which is sharp
in contrast with the vanishing sound velocity of homogenous systems~\cite{jan, iskin-mixture2}.

When $F < 0.5$ ($F > 0.5$), making the identification that the 
number of particle (hole) pairs per lattice site is $n_B = F$ ($n_B = 1 - F$),
we obtain $U_{BB} = 2g (1 - F) a_c^3$ ($U_{BB} = 2g F a_c^3$)
as the repulsive particle (hole) boson-boson interaction such that 
the Bogoliubov relation 
$
v_B^2 = U_{BB} n_{B}/(m_B a_c^3)
$
is recovered. Here, we also identified $m_B = g/(4a_c^2 t_\uparrow t_\downarrow)$
as the mass of the bound pairs to be discussed in Sec.~\ref{sec:gaussian.tc}.
Therefore, in both low and high filling 
fraction limits, we find that $U_{BB} \approx 2g a_c^3$.

\begin{figure} [htb]
\centerline{\scalebox{0.5}{\includegraphics{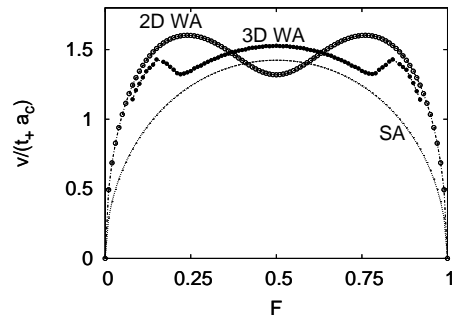} }}
\caption{\label{fig:collective}
Sound velocity $v$ (in units of $t_+ a_c$) versus $F$ in the weak attraction
(WA) limit for $g = 2t_+$ in two (hollow squares) and (solid squares) in three dimensions, 
and in the strong attraction (SA) limit for $g = 20t_+$ (dotted line) 
in both two and three dimensions.
Here, $t = t_\uparrow = t_\downarrow$ and $P = 0$.
}
\end{figure}

In Fig.~\ref{fig:collective}, we show the sound velocity $v$ for two and three dimensional
lattices as a function of $F$ when $t = t_\uparrow = t_\downarrow$ and $ P = 0$.
In the BCS limit shown for $g = 2t_+$, $v$ is very different for two and three dimensional lattices
due to van Hove singularities present in their density of states~\cite{sofo, micnas}.
However, in the BEC limit shown for $g = 20t_+$, $v$ saturates to a finite value, 
which is identical for both two and three dimensional lattices and reproduces the
analytical results discussed above.

Before concluding this section, we discuss the fraction of condensed
pairs to understand further the saturation of sound velocity in the BEC limit.
In the absence of a population imbalance ($P = 0$), 
the number of condensed pairs per lattice site is given by~\cite{leggett-He}
\begin{equation}
n_c = \frac{1}{M} \sum_{\mathbf{k}} \frac{|\Delta_\mathbf{k}|^2}{4E_{\mathbf{k},+}^2} 
\tanh^2\left(\frac{\beta E_{\mathbf{k},+}}{2}\right),
\end{equation}
where
$
\phi(\mathbf{k}) = \Delta_{\mathbf{k}}/(2E_{\mathbf{k},+})
$
is the ground state pair wave function. 
At zero temperature, while the number of condensed pairs
$
n_c = \pi {\cal N}(0) |\Delta_0|/4
$
is very small in comparison to the total number of fermions
in the BCS limit, it increases as a function of 
interaction strength and saturates in the BEC limit to
$
n_c = F(1-F),
$
leading to the number of condensed bosons per lattice site as
$
n_{B,c} = n_B (1-n_B).
$
Here, the number of bosons per lattice site is $n_B = n/2 = F$.
Therefore, in the dilute limit when $n_B \to 0$, almost all of the bosons are 
condensed such that $n_{B,c} \to n_B$, however, 
in the dense limit when $n_B \to 1$, almost all of the bosons are 
non-condensed such that $n_{B,c} \ll n_B$.
Making the identification that the number of condensed bosons is $n_{B,c} = F(1-F)$, 
we obtain $U_{BB} = 2ga_c^3$ as the repulsive particle (hole) boson-boson 
interaction such that the Bogoliubov relation
$
v_B^2 = U_{BB} n_{B,c}/(m_B a_c^3)
$
is recovered.

Having discussed Gaussian fluctuations at zero temperature, we analyze next
Gaussian fluctuations near the critical temperature and derive the 
time-dependent Ginzburg-Landau (TDGL) functional.

\subsection{Gaussian Fluctuations \\ Near the Critical Temperature}
\label{sec:gaussian.tc}

In this section, we present the results of finite temperature Gaussian fluctuations near the 
superfluid critical temperature $T_c$. We define the field $\Lambda(x)$ to be the fluctuation around the order
parameter saddle point value $|\Delta_0| = 0$, and use the same procedure 
as in the homogeneous case~\cite{carlos, iskin-mixture2}, to obtain the TDGL functional
\begin{equation}
\label{eqn:exp.Tc}
\left[ a + b|\Lambda(x)|^2 - \sum_{<i,j>} \frac{c_{i,j}}{2}\nabla^2 - id\frac{\partial}{\partial t} \right] \Lambda(x) = 0,
\end{equation}
in the position and time $x = (\mathbf{r},t)$ representation.
The expansion coefficients $a, b, c_{i,j}$ and $d$ are given in App.~\ref{sec:app.b}.
Notice that $c_{i,j} = c \delta_{i,j}$ is isotropic for the s-wave interactions 
considered in this manuscript, where $\delta_{i,j}$ is the Kronecker delta.
Next, we discuss analytically the TDGL functional in the weak and strong 
attraction limits.

In the weak attraction (BCS) limit when $0 < \widetilde{\mu}_\sigma < 4{\cal D}t_\sigma$, 
it is difficult to derive analytical expressions in the
presence of population imbalance, and thus we limit our discussion 
only to the $P = 0$ limit.
In this case, we obtain 
$a = {\cal N}(0) \ln(T/T_c)$ for the coefficient of the linear term in $\Lambda (x)$,
$b = 7{\cal N}(0)\zeta(3) / (8\pi^2 T_c^2)$ for the cofficient of the cubic term,
$c = 7{\cal N}_t(0)\zeta(3) / (4{\cal D}\pi^2 T_c^2)$ for the coefficent of the operator $\nabla^2$, and
$d = {\cal N}(0)(2{\cal D}t_+ - \widetilde{\mu}_+)/[2\widetilde{\mu}_+(4{\cal D}t_+ - \widetilde{\mu}_+)] 
+ i\pi{\cal N}(0)/(8T_c)]$ for the coefficient of the operator ${\partial}/{\partial t}$.
Here, 
$
{\cal N}_t(x) = \sum_{\mathbf{k}, i} [(\nabla_i \xi_{\mathbf{k},\uparrow})^2 + (\nabla_i \xi_{\mathbf{k},\downarrow})^2] 
\delta(\xi_{\mathbf{k},+} - x)
$
represents an effective `kinetic' density of states,
and the parameter
$
T_c = (2/\pi) [(2{\cal D}t_+)^2 - (\widetilde{\mu}_+ - 2{\cal D}t_+)^2]^{1/2} \exp[\gamma -1/(g {\cal N}(0))]
$
is the BCS critical temperature where $\gamma = 0.577$ is the Euler's constant, 
$
\widetilde{\mu}_+ = F/{\cal N}(0),
$
and $\zeta(x)$ is the Zeta function. Notice that $T_c$ is quite small for weak attractions, but increases
with growing values of $g$. In addition, notice that the imaginary part of $d$ is large, indicating 
that Cooper pairs have a finite lifetime and that they decay into the continuum of two-particle states.

In the strong attraction (BEC) limit when
$|\widetilde{\mu}_+| \approx |\epsilon_b|(1 - p_e)/2 \gg 2{\cal D}t_+$, we obtain
$a = a_1 + a_2 = -[2\widetilde{\mu}_+ - \epsilon_b(1-p_e)]/[g^2(1 - p_e)] + p_e/[g(1-p_e)]$
for the coefficient of the linear term in $\Lambda (x)$,
$b = b_1 + b_2 = 2/[g^3(1-p_e)^2] - (\partial p_e/\partial \widetilde{\mu}_e)/[g^2(1-p_e)]$
for the cofficient of the cubic term,
$c = 4a_c^2 t_\uparrow t_\downarrow/[g^3(1-p_e)^2]$ for the coefficent of the operator $\nabla^2$, and
$d = 1/[g^2(1-p_e)]$ for the coefficient of the operator ${\partial}/{\partial t}$.
Here, $\epsilon_b = - g$ is the binding energy,
$e$ ($-e$) labels the excess (non-excess) type of fermions, and 
$p_e = | n_{\uparrow} - n_{\downarrow}| $ is the number of unpaired fermions per lattice site.
Notice that the imaginary part of $d$ vanishes in this limit, reflecting the presence of long 
lived bound states.

Through the rescaling $\Psi(x) = \sqrt{d}\Lambda(x)$,
we obtain the equation of motion for a mixture of bound pairs (molecular bosons)
and unpaired (excess) fermions
\begin{eqnarray}
-\mu_B \Psi(x) &+& \left[U_{BB}|\Psi(x)|^2 + U_{BF} p_e(x) \right] \Psi(x) \nonumber \\ 
&-& \frac{\nabla^2 \Psi(x)}{2m_B}  - i\frac{\partial \Psi(x)}{\partial t} = 0,
\label{eqn:GL-BEC}
\end{eqnarray}
with boson chemical potential 
$
\mu_B = - a_1/d = 2\widetilde{\mu}_+ - \epsilon_b(1-p_e),
$
mass 
$
m_B = d/c = g/(4a_c^2 t_\uparrow t_\downarrow),
$
and repulsive boson-boson
$
U_{BB} = b_1 a_c^3/d^2 = 2g a_c^3
$
and boson-fermion
$
U_{BF} = a_2 a_c^3 /(d p_e) = g a_c^3
$
interactions.
Notice that, the repulsion $U_{BB}$ between bound pairs is two times larger than 
the repulsion $U_{BF}$ between a bound pair and a fermion, reflecting the Pauli
exclusion principle. Furthermore, $U_{BB}$ and $U_{BF}$ 
are strongly repulsive due to the important role played by the Pauli exclusion principle 
in the lattice~\cite{iskin-lattice}, in contrast to the 
homogenous case~\cite{carlos, pieri, iskin-mixture, iskin-mixture2} 
where $U_{BB}$ and $U_{BF}$ are weakly repulsive.

This procedure also yields
$p_e (x) = [a_2/d + b_2|\Psi(x)|^2/d^2]/U_{BF}$,
which is the spatial density of unpaired fermions
\begin{equation} 
p_e (x) = p_e - g a_c^3(\partial p_e/\partial \mu_e) (1-p_e) |\Psi(x)|^2 \ge 0.
\end{equation}
The critical temperature in the case of zero population imbalance can be obtained
directly from the effective boson mass $m_B$, using the standard BEC condition
$T_{BEC} = 2 \pi \left[ n_B \zeta (3/2) \right]^{2/3}/m_B $ leading to 
$T_c \propto t_{\uparrow} t_{\downarrow} / g$, which decreases with growing attraction $g$.
Here, $\zeta(x)$ is the zeta function.
Notice that $T_c$ in the lattice case vanishes for $g \to \infty$ unlike the homogeneous case 
which saturates at $T_c \approx 0.218 \epsilon_F$, where $\epsilon_F$ is the Fermi energy~\cite{nsr, carlos}.
The decrease of $T_c$ with increasing $g$ in the BEC regime, combined with the 
increase of $T_c$ with increasing $g$ in the BCS regime leads to a maximum 
in between as already noted in the literature~\cite{nsr, micnas}. 

Further insight into the differences between the homogeneous and the lattice systems 
can be gained by comparing the Ginzburg-Landau coherence length  
$
\xi (T) = [c/(2 a)]^{1/2},
$
which in the vicinity of $T_c$ becomes
$
\xi (T) = \xi_{GL} [T_c/(T_c-T)]^{1/2}.
$
The prefactor of the coherence length is
\begin{equation}
\xi_{GL} = \left[ \frac{c}{2T_c} \left( \frac{\partial a}{\partial T} \right)^{-1} \right]_{T=T_c}^{1/2},
\end{equation}
and is evaluated at $T_c$.
While 
$
\xi_{GL} = [7{\cal N}_t(0)\zeta(3) / (8{\cal D}\pi^2 T_c^2)]^{1/2}
$
is large in the weak attraction limit in comparison to the lattice spacing $a_c$,
$
\xi_{GL} = a_c(t_\uparrow t_\downarrow/F)^{1/2}/g
$ 
is small and phase coherence is lost gradually as $g$ increases in the strong 
attraction limit when $g/t_+ \gg 1$. 
This latter result is in sharp contrast with the homogenous case~\cite{carlos}, 
where $\xi_{GL}$ is large compared to interparticle spacing in both BCS and BEC limits, 
and it has a minimum near $\widetilde{\mu}_+ \approx 0$~\cite{carlos}. However,
$\xi_{GL}$ depends inversely on the square-root of the density in 
both homogenous and lattice systems.

Since the boson-boson and boson-fermion interactions are strongly repulsive 
in the BEC limit (strong attraction regime for fermions) due to the 
important role played by the Pauli exclusion principle in the lattice~\cite{iskin-lattice},
it is necessary to investigate this further, as discussed next.

\section{Strong Attraction (Molecular) Limit}
\label{sec:molecular}

In the strong attraction limit, the system can be described by an action containing
molecular bosons and excess fermions~\cite{pieri, iskin-mixture, iskin-mixture2}. 
The existence of an optical lattice produces different 
physics from the homogeneous case in the strong attraction limit, because of the 
strong repulsive interaction between molecular bosons and excess fermions~\cite{iskin-lattice}.
The main effect of these strong effective interactions is the emergence of insulating
phases from the resulting Bose-Fermi mixture of molecular bosons and excess fermions in
a lattice. 

\subsection{Effective Lattice Bose-Fermi Action}
\label{sec:BFaction}

In the limit of strong attractions between
fermions $(g/t_{+} \gg 1)$, we obtain an effective Bose-Fermi lattice action~\cite{iskin-lattice}
\begin{equation}
S_{BF}^{\rm eff} =  \int_0^{\beta} d\tau  \left[ \sum_{i} ( f_i^\dagger \partial_{\tau} f_i + 
b_i^\dagger \partial_{\tau} b_i)  +  {H}_{BF}^{\rm eff} \right], 
\end{equation}
where
$
{H}_{BF}^{\rm eff} = K_{F} + K_{B} + H_{BF} + H_{BB}.
$
Here, the kinetic part of the excess fermions is 
\begin{equation}
\label{eqn:kf}
K_{F} = -\mu_F \sum_i  f_i^\dagger f_i 
- t_F \sum_{\langle i,j \rangle} f_i^\dagger f_j ;
\end{equation}
the kinetic part of the molecular bosons is 
\begin{equation}
\label{eqn:kb}
K_B = - \mu_{B} \sum_i  b_i^\dagger b_i 
- t_B \sum_{\langle i,j \rangle} b_i^\dagger b_j;
\end{equation}
the interaction between molecular bosons and excess fermions is 
\begin{equation}
\label{eqn:hbf}
H_{BF} =  U_{BF} \sum_{i} f_i^\dagger f_i b_i^\dagger b_i + 
V_{BF} \sum_{\langle i, j \rangle} f_i^\dagger f_i b_j^\dagger b_j;
\end{equation}
and the interaction between two molecular bosons is 
\begin{equation}
\label{eqn:hbb}
H_{BB}  = U_{BB} \sum_{i} b_i^\dagger b_i b_i^\dagger b_i +
V_{BB} \sum_{\langle i, j \rangle} b_i^\dagger b_i b_j^\dagger b_j. 
\end{equation}

The total number of fermions is fixed by the constraint
$n = 2n_B + p_e$, where $n_B = N_B/M$ is the number of bosons per lattice site.
The important parameters of this effective Hamiltonian are the excess fermion transfer energy $t_F = t_e$, 
the molecular boson transfer energy $t_B = 2 t_{\uparrow} t_{\downarrow}/g$,
the boson-fermion effective repulsion $U_{BF} = g$ and the boson-boson
effective repulsion $U_{BB} = 2g$.
Notice that, on-site interactions $U_{BB}$ and $U_{BF}$ become infinite (hard-core)
when $g \to \infty$ as a manifestation of the Pauli exclusion principle.
In addition, there are weak and repulsive nearest neighbor boson-boson 
$V_{BB} \propto (t_{\uparrow}^2 + t_{\downarrow}^2)/g$ and
boson-fermion $V_{BF} \propto t_e^2/g$ interactions. 
These repulsive interactions in optical lattices lead to several insulating phases,
depending on fermion filling fractions, as discussed next.

\subsection{Emergence of Insulating Phases}
\label{sec:insulating}

When the fermion attraction is sufficiently strong, the lines BC and CD must 
describe insulators, as molecular bosons and excess fermions are strongly 
repulsive.
In the following analysis, we discuss only two high symmetry cases:
(a) $n_\uparrow = n_\downarrow = 1$; and
(b) $n_\uparrow = 1$ and $n_\downarrow = 1/2$, or  $n_\uparrow  = 1/2$ and $n_\downarrow = 1$.

\begin{figure} [htb]
\centerline{\scalebox{0.4}{\includegraphics{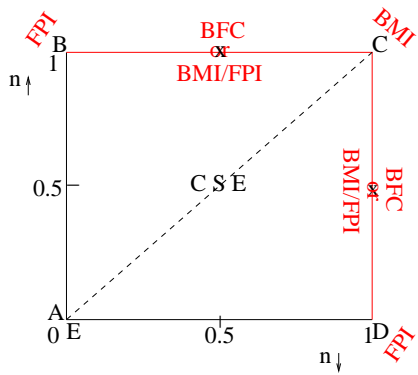}} }
\caption{\label{fig:phasei}
(Color online) $n_\uparrow$ versus $n_\downarrow$ phase diagram 
in the strong attraction limit, indicating the CSE
(superfluid), the metallic (lines AB and DE), and 
insulating (lines BC and CD) phases with special points Fermi-Pauli insulator (FPI), 
Bose-Fermi checkerboard (BFC), Bose-Mott insulator (BMI), and BMI/FPI phase separation.
}
\end{figure}

Case (a) is indicated as point C in Figs.~\ref{fig:lattice.tr.1},~\ref{fig:lattice.tr.6.67} 
and~\ref{fig:phasei} where $n_\uparrow = n_\downarrow = 1$ such that $p_e = 0$.
This symmetry point is a Fermi-Pauli (band) insulator for weak attraction since 
both fermion bands are fully occupied, and a Bose-Mott Insulator (BMI) 
in the strong attraction limit. 
There is exactly one molecular boson (consisting of a pair of $\uparrow$ and 
$\downarrow$ fermions) at each lattice site, which has a strong repulsive on-site 
interaction with any additional molecular boson due to the Pauli exclusion principle.
Notice that this case is very similar to the atomic BMI transition observed 
with one-species atomic Bose systems~\cite{bloch-review}.

\begin{figure} [htb]
\centerline{\scalebox{0.45}{\includegraphics{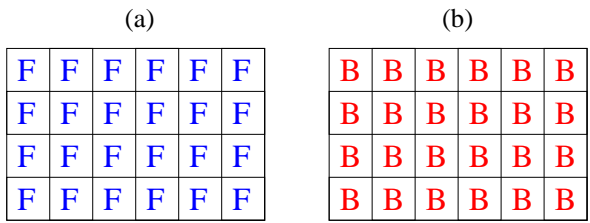}} }
\caption{\label{fig:lattice.bmi}
(Color online) Schematic diagrams for the 
(a) Fermi-Pauli insulator (FPI), and 
(b) Bose-Mott insulator (BMI) phases.
}
\end{figure}

In case (a), $H_{BF}^{\rm eff}$ reduces to a molecular Bose-Hubbard Hamiltonian with the 
molecular Bose filling fraction $n_B = n/2 = F$, thus leading to 
a molecular BMI when $n_B = 1$ beyond a critical value of $U_{BB}$.
A schematic diagram of this phase is shown in Fig.~\ref{fig:lattice.bmi}(b).
The critical value $U_{BB}^c$ needed 
to attain the BMI phase can be estimated using the approach of Ref.~\cite{stoof-2001}
leading to $U_{BB}^c = 3 (3 + \sqrt{8}) t_B$, which in terms of the underlying
fermion parameters leads to $g_c = 4.18 \sqrt{t_{\uparrow} t_{\downarrow}}$
for the critical fermion interaction. This value of $g_c$ is just a lower bound
of the superfluid-to-insulator (SI) transition, since $H_{BF}^{\rm eff}$ is only valid in 
the $g \gg t_+$ limit. Some signatures of this SI transition at $g_c$ have been observed 
in recent experiments~\cite{mit-lattice}. 

We would like to make some remarks on the validity $H_{BF}^{\rm eff}$ presented above, which intrinsically
assumes that the molecular bosons are small in comparison to the lattice spacing. A measure
of the `smallness' of these molecular bosons is the average size of the fermion 
pairs defined (for $p_e = 0$) by
\begin{equation}
\xi_{pair} = \left[ \frac{\langle \phi(\mathbf{k})|r^2|\phi(\mathbf{k}) \rangle} 
{\langle \phi(\mathbf{k})|\phi(\mathbf{k}) \rangle} \right]^{1/2},
\end{equation}
where 
$
\phi(\mathbf{k}) = \Delta_{\mathbf{k}}/(2E_{\mathbf{k},+})
$
is the ground state pair wave function.
While 
$
\xi_{pair} = v_F/(4|\Delta_0|)
$
is large in the weak attraction limit in comparison to the lattice spacing $a_c$,
$
\xi_{pair} = 2 \sqrt{2{\cal D}} a_c |1 - 2F| t_+/g
$ 
in the strong attraction limit when $g/t_+ \gg 1$.
These expressions are valid for low and high filling fractions, $F \to 0$ and $F \to 1$, respectively.
At $g = g_c$, we obtain $\xi_{pair} \approx 1.17 a_c$ for $t_\uparrow = t_\downarrow$ 
and $\xi_{pair}/a_c \approx 1.74 a_c$ for $t_\uparrow = 0.15 t_{\downarrow}$. 
Notice that larger values of $g$ lead to molecular boson sizes 
that are smaller than the lattice spacing, thus validating the 
effective Bose-Fermi description derived above for $g \gg t_{+}$ to values of $g$ close to $g_c$.
In the lattice case, the results above suggest that even if the size of the bound pairs
decreases as $t_+ a_c/g$ with increasing $g$,
the Pauli pressure prohibits two bound pairs or a bound pair and an excess fermion
to occupy the same lattice site, in the single-band description discussed here. 
This indicates that the bound pairs do not loose their fermionic nature and can not
be thought as structureless in lattices, which should be contrasted
with the homogenous case where bound pairs become structureless with increasing attraction.

\begin{figure} [htb]
\centerline{\scalebox{0.45}{\includegraphics{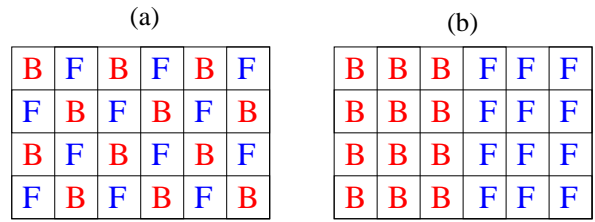}} }
\caption{\label{fig:lattice.bfc}
(Color online) Schematic diagrams for the 
(a) Bose-Fermi checkerboard (BFC) phase, and 
(b) Bose-Mott insulator (BMI) and Fermi-Pauli insulator (FPI) phase separation.
}
\end{figure}

Case (b) is indicated as crosses in Figs.~\ref{fig:lattice.tr.1},~\ref{fig:lattice.tr.6.67} 
and~\ref{fig:phasei} at points $n_\uparrow = 1, n_\downarrow = 1/2$
or $n_\uparrow = 1/2, n_\downarrow = 1$, where the molecular boson filling fraction 
$n_B = 1/2$ and the excess fermion filling fraction is $p_e = 1/2$.
At these high symmetry points, molecular bosons and excess fermions tend
to segregate, either producing a domain wall type of phase separation between a molecular 
Bose-Mott insulator (BMI) region and a Fermi-Pauli insulator (FPI) region 
or a checkerboard phase of alternating molecular bosons and excess fermions (BFC)
depending on the ratio $V_{BB}/V_{BF}$. 
A schematic diagram of these two phases is shown in Fig.~\ref{fig:lattice.bfc}.

The checkerboard phase shown in Fig~\ref{fig:lattice.bfc}(a) is favored 
when $V_{BB} > 2V_{BF}$. At the current level of approximation,
we find that when $t_\uparrow = t_\downarrow$ phase separation is 
always favored, however when $\uparrow$ ($\downarrow$) fermions are in excess the checkerboard phase 
is favored when $t_\downarrow > \sqrt{3} t_\uparrow$ 
$(t_\downarrow < t_\uparrow/\sqrt{3})$.
Therefore, phase separation and checkerboard phases are achievable 
if the tunneling ratio $\eta$ can be controlled experimentally in optical
lattices. Notice that this checkerboard phase present in the lattice case 
is completely absent in homogeneous or harmonically trapped 
systems~\cite{pao, sheehy, iskin-mixture, iskin-mixture2}.

Thus, the strong attraction limit in optical lattices brings additional physics 
not captured at the saddle point, and not present in homogenous or purely harmonically 
trapped systems. Having discussed the strong attraction (molecular) limit and its possible
phases, we comment next on their experimental detection.

\section{Detection of Insulating Phases}
\label{sec:detection}

It may be very difficult to detect all the superfluid and insulating phases 
proposed {\it in situ}, and thus one may have to turn off the optical lattice 
to perform time-of-flight measurements.
The phases proposed here in the molecular limit could be detected either
via the measurement of momentum distribution, density-density correlations
or density Fourier transform. For instance, the Fermi-Pauli insulator (FPI), Bose-Mott insulator
(BMI), Bose-Fermi checkerboard (BFC), and phase separated BMI/FPI do not exhibit
phase coherence in their momentum distribution, while the superfluid phase does.
Therefore, through the momentum distribution measurements, one should be able to 
differentiate between superfluid and insulating phases, which has been possible for
Bose systems in optical lattices~\cite{bloch-review}. 
However, in order to detect clearly the different insulating phases, 
it may also be necessary to explore the Fourier transforms of the densities,
or of the density-density correlation functions.

The relevant density-density correlation functions to characterize the insulating phases are 
$
C_{F,F} (\mathbf{r}, \mathbf{r'}) = \langle n_F (\mathbf{r}) n_F (\mathbf{r'}) \rangle
$ 
which correlates the excess fermions, 
$
C_{B,B} (\mathbf{r}, \mathbf{r'} ) = \langle n_B (\mathbf{r}) n_B (\mathbf{r'}) \rangle
$
which correlates the molecular bosons, and 
$
C_{B,F} (\mathbf{r}, \mathbf{r'}) = \langle n_B (\mathbf{r}) n_F (\mathbf{r'}) \rangle
$
which correlates the molecular bosons and excess fermions. Defining the relative 
$\mathbf{R} = \mathbf{r} - \mathbf{r'}$ and the sum $\mathbf{R}_c = (\mathbf{r} + \mathbf{r'})/2 $ 
of the coordinates, we can write the correlation functions as 
$
C_{\alpha,\beta} (\mathbf{R}, \mathbf{R}_c) = \langle n_\alpha ( \mathbf{R}_c + \mathbf{R}/2 ) 
n_\beta ( \mathbf{R}_c - \mathbf{R}/2 ) \rangle,
$ 
where $\alpha$ and $\beta $ are $\{F,B\}$. This leads to the Fourier transform
$\widetilde C_{\alpha,\beta} (\mathbf{q}, \mathbf{q}_c)$, where $\mathbf{q}$ is the relative 
and $\mathbf{q}_c$ is the sum of the momentums. For a translationally invariant system 
considered here, it is sufficient to look only to $\mathbf{q}_c = \mathbf{0}$, and define
$
\widetilde C_{\alpha,\beta} (\mathbf{q}) \equiv \widetilde C_{\alpha,\beta} (\mathbf{q}, \mathbf{0}).
$
Therefore, for the FPI, BMI and BFC insulating phases, correlation functions have the generic form 
$
\widetilde C_{\alpha,\beta} (\mathbf{q}) = 
\sum_{i,j} \exp [-i \mathbf{q} \cdot (\mathbf{r_i}_{,\alpha} - \mathbf{r_j}_{,\beta}) ],
$
where $\mathbf{r_i}_{,\alpha}$ and $\mathbf{r_j}_{,\beta} $ are the locations of
$\alpha$ and $\beta$ particles.

For instance, in the case of a two-dimensional system, the FPI phase [shown in Fig.~\ref{fig:lattice.bmi}(a)]
has a strong peak in the correlation function $\widetilde C_{F,F} (\mathbf{q})$ at $\mathbf{q}  = (0,0)$.
Similarly, the BMI phase [shown in Fig.~\ref{fig:lattice.bmi}(b)]
has a strong peak in the correlation function $\widetilde C_{B,B} (\mathbf{q})$ at $\mathbf{q}  = (0,0)$. 
However, the BFC phase [shown in Fig.~\ref{fig:lattice.bfc}(a)]
has strong peaks in the correlation functions $\widetilde C_{F,F} (\mathbf{q})$ and
$\widetilde C_{B,B} ( \mathbf{q} )$ both at $\mathbf{q}  = (0,0)$ and $\mathbf{q}  = \pm (\pi,\pm \pi)$,
while it has strong peaks in the correlation function $\widetilde C_{B,F} (\mathbf{q})$  
at $\mathbf{q} = (0,0)$. 

The insulating FPI, BMI, and BFC phases are modified 
in the presence of an overall harmonic trapping potential,
since the molecular bosons experience the harmonic potential 
$
V_B (r) = (m_{\uparrow} \omega_{\uparrow}^2 + m_{\downarrow} \omega_{\downarrow}^2) r^2/2,
$
and the excess fermions of type $\sigma$ experience a different harmonic potential 
$
V_{F, \sigma} (r) = m_{\sigma} \omega_{\sigma}^2 r^2/2.
$
Provided that the harmonic potentials are not too strong, 
the characteristic peak locations of the correlation functions remain essentially the same, 
but additional broadening due to the inhomogeneous nature of the trapping potential occurs. 
However, when the harmonic potentials are sufficiently large, important modifications of these
phases are present, including the emergence of complex shell structures.

Lastly, the BMI/FPI phase separated state is strongly modified even in the presence of
weak trapping potentials. The single domain wall density structure [shown in
Fig.~\ref{fig:lattice.bfc}(b)] turns into a new density profile with a region where 
the bosons tend to stay at the center of the trapping potential due in part
to their tighter confinement, and the excess fermions are pushed out of the center by the strong Bose-Fermi repulsion.
The fermions then surround the bosons, and standard {\it in situ} radio-frequency measurements 
can detect the existence of the BMI/FPI phase separated state, as done recently
with population imbalanced fermion mixtures~\cite{mit, rice, mit-2, rice-2}. 

Having commented briefly on the experimental detection of the superfluid and insulating phases, 
we present our conclusions next.

\section{Conclusions}
\label{sec:conclusions}

Using an attractive Fermi-Hubbard Hamiltonian to describe mixtures of one- or two-species
of atoms in optical lattices, we obtained the ground state phase diagram of Fermi-Fermi mixtures 
containing normal, phase-separated and coexisting superfluid/excess fermions, and insulating regions.
We discussed the cases of balanced and imbalanced populations of Fermi-Fermi mixtures
such as $^6$Li or $^{40}$K only; and mixtures of $^6$Li and $^{40}$K; $^{6}$Li and $^{87}$Sr; or $^{40}$K and $^{87}$Sr.
We showed that population imbalanced Fermi-Fermi mixtures 
reduce to strongly interacting (repulsive) Bose-Fermi mixtures in the molecular limit, 
in sharp contrast to homogenous systems where the resulting Bose-Fermi mixtures
are weakly interacting. This result is a direct manifestation of the Pauli exclusion principle in the lattice case, 
since each Bose molecule consists of two fermions, and more than one identical 
fermion on the same lattice site is not allowed, within a single-band description. 
This effect together with the Hartree energy shift lead to a filling dependent condensate
fraction and to sound velocities which do not approach zero, in contrast to the homogenous case.

Furthermore, we showed that several insulating phases appear in the strong attraction limit depending 
on filling fraction and population imbalance. For instance, we found 
a molecular Bose-Mott insulator (superfluid) when the molecular filling 
fraction is equal to (less than) one 
for a population balanced system where the fermion filling fractions
are identical. When the filling fraction of one type of fermion is one and the filling fraction of the other
is one-half (corresponding to molecular boson and excess fermion filling fractions of one-half),
we also found either a phase-separated state consisting of a Fermi-Pauli insulator (FPI) 
of the excess fermions and a molecular Bose-Mott insulator (BMI) or a Bose-Fermi checkerboard 
(BFC) phase depending on the tunneling anisotropy ratio.

All of these additional phases and the possibility of observing more exotic superfluid 
phases in optical lattices with p-wave order parameters~\cite{iskin-prb} 
make the physics of Fermi-Fermi mixtures much richer than those of atomic bosons
or Bose-Fermi mixtures in optical lattices, and of harmonically trapped fermions.
Lastly, the molecular BMI phase discussed here has been 
preliminarily observed in a very recent experiment~\cite{mit-lattice}, 
opening up the experimental exploration of the rich phase diagram of  
fermion mixtures in optical lattices in the near future.

We thank National Science Foundation (DMR-0709584) for support.

\appendix

\section{Expansion Coefficients at Zero Temperature}
\label{sec:app.a}

In this Appendix, we perform a small $\mathbf{q}$ and $w$ 
expansion of the effective action at zero temperature ($T = 0$)~\cite{iskin-mixture2}.
In the amplitude-phase basis, we obtain the expansion coefficients necessary 
to calculate the collective modes, as discussed in Section~\ref{sec:gaussian.zero}. 
We calculate the coefficients only for the 
case of s-wave pairing with zero population imbalance $P = 0$, as extra care is needed when 
$P \ne 0$ due to Landau damping. In the long-wavelength $(|\mathbf{q}| \to 0)$, 
and low-frequency $(w \to 0)$ limits the condition
$
\{ w , \epsilon_{\mathbf{q},+} \} \ll \min \{ 2E_{\mathbf{k},+} \}
$
is used. 

The coefficients necessary to obtain the diagonal amplitude-amplitude matrix element are
\begin{equation}
A = \frac{1}{M} \sum_{\mathbf{k}} \frac{|\Delta_0|^2} {2E_{\mathbf{k},+}^3} 
\end{equation}
corresponding to the $(\mathbf{q} = 0, w = 0)$ term,
\begin{eqnarray}
C &=& \frac{a_c^2}{M} \sum_{\mathbf{k}} \Big\lbrace
\frac{E_{\mathbf{k},+}^2 - 3|\Delta_0|^2}{4 E_{\mathbf{k},+}^5} \xi_{\mathbf{k},+} t_+ \cos(k_xa_c) \nonumber \\
&-& \left[\left( E_{\mathbf{k},+}^2 - 10|\Delta_0|^2
+ \frac{10 |\Delta_0|^4}{E_{\mathbf{k},+}^2} \right) t_+^2 \right. \nonumber \\
&+& \left. 
(E_{\mathbf{k},+}^2 - |\Delta_0|^2) t_-^2
\right] \frac{\sin^2(k_xa_c)}{2 E_{\mathbf{k},+}^5}
\Big\rbrace
\end{eqnarray}
corresponding to the $|\mathbf{q}|^2$ term, and
\begin{equation}
D = \frac{1}{M} \sum_{\mathbf{k}} \frac{E_{\mathbf{k},+}^2 - |\Delta_0|^2} {8E_{\mathbf{k},+}^5}
\end{equation}
corresponding to the $w^2$ term.

The coefficients necessary to obtain the diagonal phase-phase matrix element are
\begin{eqnarray}
Q &=& \frac{a_c^2}{M} \sum_{\mathbf{k}} \Big\lbrace
\frac{\xi_{\mathbf{k},+} t_+} {4 E_{\mathbf{k},+}^3} \cos(k_x a_c)
- \left[ (E_{\mathbf{k},+}^2 - 3|\Delta_0|^2) t_+^2 \right. \nonumber \\
&+& \left. E_{\mathbf{k},+}^2 t_-^2 \right] 
\frac{\sin^2(k_xa_c)} {2 E_{\mathbf{k},+}^5} 
\Big\rbrace
\end{eqnarray}
corresponding to the $|\mathbf{q}|^2$ term, and
\begin{equation}
R = \frac{1}{M} \sum_{\mathbf{k}} \frac{1} {8E_{\mathbf{k},+}^3}
\end{equation}
corresponding to the $w^2$ term.

The coefficient necessary to obtain the off-diagonal matrix element is
\begin{equation}
B = \frac{1}{M} \sum_{\mathbf{k}} \frac{\xi_{\mathbf{k},+}} {4E_{\mathbf{k},+}^3}
\end{equation}
corresponding to the $w$ term.
These coefficients can be evaluated analytically in the BCS and BEC limits, and
are given in Sec.~\ref{sec:gaussian.zero}.

\section{Expansion Coefficients Near the Critical Temperature}
\label{sec:app.b}

In this Appendix, we derive the coefficients $a, b, c_{i,j}$ and $d$ of the time
dependent Ginzburg-Landau theory described in Section~\ref{sec:gaussian.tc}. 
We perform a small $\mathbf{q}$ and $w$ expansion of the effective action 
near the critical temperature ($T \approx T_c$), where we assume that the 
fluctuation field $\Lambda(\mathbf{r}, t)$ is a slowly varying function of 
position $\mathbf{r}$ and time $t$~\cite{iskin-mixture2}.

The zeroth order coefficient is given by
\begin{equation}
a = \frac{1}{g} - \frac{1}{M}\sum_{\mathbf{k}} \frac{X_{\mathbf{k},+}} {2\xi_{\mathbf{k},+}}
|\Gamma_s (\mathbf{k})|^2
\end{equation}
where 
$
X_{\mathbf{k},\pm} = (X_{\mathbf{k},\uparrow} \pm X_{\mathbf{k},\downarrow})/2
$
and
$
X_{\mathbf{k},\sigma} = \tanh(\beta\xi_{\mathbf{k},\sigma}/2).
$

The second order coefficient is given by 
\begin{eqnarray}
c_{i,j} &=& \frac{a_c^2}{M} \sum_{\mathbf{k}} \Big\lbrace
	\left( X_{\mathbf{k},\uparrow} Y_{\mathbf{k},\uparrow} t_\uparrow^2 
	+ X_{\mathbf{k},\downarrow} Y_{\mathbf{k},\downarrow} t_\downarrow^2 \right)
	\sin(k_ia_c) \nonumber \\
	&&\sin(k_ja_c) \frac{\beta^2} {8\xi_{\mathbf{k},+}} 
	+ \left[ \frac{4 t_- \sin(k_ia_c) \sin(k_ja_c) C_-}{\xi_{\mathbf{k},+}} \right. \nonumber \\
	&&\left. - 2\delta_{i,j} \cos(k_ia_c) C_+ \right] \frac{\beta}{8 \xi_{\mathbf{k},+}}
  + \left[2t_+ \cos(k_ia_c) \delta_{i,j} \right. \nonumber \\
  && \left.- \frac{4t_-^2 \sin(k_ia_c) \sin(k_ja_c)}{\xi_{\mathbf{k},+}}\right] 
	\frac{X_{\mathbf{k},+}} {4\xi_{\mathbf{k},+}^2}
\Big\rbrace |\Gamma_s (\mathbf{k})|^2,
\end{eqnarray}
where 
$
C_\pm = (Y_{\mathbf{k},\uparrow} t_\uparrow \pm Y_{\mathbf{k},\downarrow} t_\downarrow) / 2,
$
$
$ 
and 
$
Y_{\mathbf{k},\sigma} = {\rm sech}^2(\beta\xi_{\mathbf{k},\sigma}/2).
$
Here, $\delta_{i,j}$ is the Kronecker delta.

The fourth order coefficient is given by
\begin{eqnarray}
b = \frac{1}{M}\sum_{\mathbf{k}}
\left(
\frac{X_{\mathbf{k},+}}{4\xi_{\mathbf{k},+}^3}
- \frac{\beta Y_{\mathbf{k},+}}{8\xi_{\mathbf{k},+}^2} 
\right) |\Gamma_s (\mathbf{k})|^4.
\end{eqnarray}

The time-dependent coefficient has real and imaginary parts, and for the s-wave case is given by
\begin{eqnarray}
d = \lim_{w \to 0} \sum_{\mathbf{k}}
\frac{X_{\mathbf{k},+}}{M}
\left[ \frac{1}{4 \xi_{\mathbf{k},+}^2} + i \frac{\pi}{w} \delta(2\xi_{\mathbf{k},+} - w)
\right] |\Gamma_s (\mathbf{k})|^2
\end{eqnarray}
where $\delta (x)$ is the delta function.
These coefficients can be evaluated analytically in the BCS and BEC limits, and
are given in Sec.~\ref{sec:gaussian.tc}.

\end{document}